\documentstyle[sprocl,epsfig]{article}

\def\be{\begin{equation}}
\def\ee{\end{equation}}
\def\bea{\begin{eqnarray}}
\def\eea{\end{eqnarray}}

\def \beq {\begin{equation}}
\def \eeq {\end{equation}}
\def \beqa {\begin{eqnarray}}
\def \eeqa {\end{eqnarray}}

\def \p {\pi}
\def \m {\mu}

\begin{document}

\begin{flushright}
YITP-SB-00-87\\
BNL-HET-00/45
\end{flushright}

\vbox{\vskip 0.25 true in}

\title{PERTURBATIVE QUANTUM FIELD THEORY\footnote{Based
on a talk presented at the {\it International Conference
on Fundamental Sciences Focusing on Mathematics
and Theoretical  Physics:  Challenges for the Twenty-First
Century}, 13-17 March, 2000,
Singapore, Singapore}}

\author{GEORGE STERMAN}

\address{C.N.\ Yang Institute for Theoretical Physics\\
State University of New York at Stony Brook\\
Stony Brook, NY 11794-3840, USA\footnote{Permanent address.}\\
and\\
Physics Department,
Brookhaven National Laboratory\\
Upton, NY 11973, USA\\
E-mail: sterman@insti.physics.sunysb.edu}


\maketitle\abstracts{This talk introduces
perturbative quantum field on a heuristic level.
It is directed at an
audience familiar with elements of quantum mechanics, but
not necessarily with high energy physics.  It
includes a discussion of the strategies behind experimental
tests of fundamental theories, and of the field theory
interpretations of these tests.}

\section*{Prologue: the Lunar Apsides}

Two and a half centuries ago, within fifty years of its author's death,
Newtonian
physics was the leading description of celestial and terrestrial motion.
Newtonian gravitation, however, was not universally accepted, even among the
great {\it savants} whose names are still familiar today. It worked as an
approximation, that much was clear, and the beauty of the Newtonian solution
to the two-body problem was indisputable.
Some perceived the theory, and its elegant inverse-square law, as
well-nigh perfect, but others felt that the law,
and its action-at-a-distance, could not be the final word. Thus,
they set out to delineate its limitations.
To these thinkers,  Newtonian
gravitation was at best a convenient accounting for appearances.

Small but cumulative changes in
planetary orbits were ideal challenges for Newtonian gravity.
  Here were what we now call
precision tests of the theory, much more demanding than the
oversimplified two-body problem. Wonderfully accurate data on
lunar motion (studied in part for its application to the
rapidly developing science of navigation) seemed a natural
place to start the search. The correspondence
on this issue of the great
mathematician and physicist Leonhard Euler, and of two of his scarcely
lesser contemporaries, Jean d'Alembert and Alexis Clairaut, illustrate a
pace of intellectual advance that we may well envy today, as well
as recognizably human generosity and jealousy, and the
eternal desire for priority. \cite{Hankins}

Between the fall of 1747 and the spring of 1749, the motion of the lunar
apsides (apogee and perigee) was a
burning question for
the Royal Academies of Paris, Berlin
and Saint Petersburg. Four milestones in the
development of its
theory are illustrated by
these communications:

\begin{itemize}

\item{}Sept.\ 1747 (Euler to Clairaut): ``\dots the forces that act on the
moon do
not follow the rule of Newton."

\item{}Nov.\ 1747 (Clairaut to French Royal Academy): [Newton's theory of
gravitation false]

\item{} June 1748 (d'Alembert to the Swiss mathematician
Cramer): ``\dots the gravitation of the moon
to the
sun will not explain [the] irregularities of its motion."

\item{} May 1749 (Clairaut to French Royal Academy): [Terms previously
neglected show Newton's law
was right all along.]

\end{itemize}

At various stages in the discussion, the possibility of a $1/r^4$ correction
to earth-moon
gravity, perhaps associated with magnetism,
was considered by these authors. Eventually,
however, the theory of lunar motion developed into the theory of the
three-body problem,
and it was here, in the influence of the sun, that an
explanation adequate to the observations was found. This was the beginning of
perturbation
theory, developed by Lagrange and Laplace, on
through to Poincar\'e and into the present day.

Euler never accepted Newtonian gravity, even as he went on to perfect a
Newtonian lunar
theory, building on Clairaut's discovery. He even arranged that Clairaut be
awarded a prize
by the Russian Royal Academy for his work. On the
other hand, relations between d'Alembert and Clairaut never
recovered from their competition on the lunar apsides.

\section{Introduction}

In our study of fundamental forces and constituents of matter, we find
ourselves following the tradition recalled above.  We are heirs to the
historic discovery of the quantum field theories that
make up the standard model.  We are learning to spin out the
consequences of these theories, while searching for a horizon beyond
which they fail, and must be replaced by a new picture of fundamental
science.

In this talk, I will describe the role of perturbative
quantum field theory in our understanding of fundamental processes in nature,
beginning
from basic concepts of quantum mechanics. We should
keep in mind that any such discussion is in the context of what I'd like to
call
``postmodern" particle physics, with its dialectic of
paired concepts, such as {\it standard} versus {\it new}, {\it effective}
versus {\it
fundamental} and {\it weak coupling} versus {\it strong coupling}. Although
usually employed in
the search for new physics, these ideas illuminate, and are
illuminated by,
quantum chromodynamics (QCD), the theory of the strong interactions in the
standard model. Even
while we are engaged in the search for physics beyond the standard model, we
are in a veritable golden age of data on the
strong interactions, in terms of
coverage and
quality. The challenge of understanding this constellation of data is, in
some ways, similar to
that which confronted Euler and his contemporaries so long ago.

The perturbative approach begins with a ``solvable'' system,
a classical or quantum mechanical Hamiltonian, $H$, whose
time-development can be quantified fully.  Perturbation theory
organizes the effects of a small modification of $H$, the
perturbation.  In celestial mechanics, this may be the influence
of a far-off ``third body'', in field theory, a small amplitude
for the quantum-mechanical production of a new particle.
At any given time, the perturbation is supposed to be a small
contribution to the energy or its expectation value; over long
periods, its influence may be profound.

I will begin with a quick introduction to perturbative quantum field theory,
and go on to discuss how perturbative techniques are
used to search for new physical phenomena. This will lead us to the technique
of separating scales, applied to ultraviolet divergences and
renormalization, on the one hand, and infrared divergences and
infrared safety on the other.  I will end with a brief discussion of how
  factorization extends the applicability of quantum field
theory in the modern experimental program.

General-interest and technical discussions of the standard
model, and of the high energy experimental program, may be
found through the web sites of the major experimental laboratories,
including CERN (www.cern.ch), DESY (www.desy.de),
Fermilab (www.fnal.gov),
and SLAC (www.slac.stanford.edu), and of the Particle
Data Group (pdg.lbl.gov).  A review of recent
(and rapidly-evolving) trends in
physics beyond the standard model is
in Ref.\ \cite{giudice}.  These
considerations serve as the leading motivation
for the endeavor of high energy physics.
The quantum-mechanical basis of the high energy tests
that they suggest is one focus of this talk.
More technical introductions
to another primary focus, the standard model example of
perturbative QCD, are in Refs.\ \cite{cteq},
\cite{tasigs} and\
\cite{tasids}.

\section{Perturbative Quantum Field Theory}

\subsection{Lagrangians and Fields}

The quantization of a field theory, particularly a gauge theory like
electrodynamics,
is rife with subtlety. Yet, the underlying principles
are straightforward enough,
and follow
the general rules of quantum mechanics. Let's ignore
the fine points, and follow a familiar path, starting with a classical
Lagrangian for
electrodynamics, and its much younger sibling,
chromodynamics. The group theory underlying the latter need not concern us
too much.
For now, we observe that the Lagrangian for electrodynamics
may be represented as
\beq
{\rm L}_{\rm EM} = K_{\rm electron} - \int d^3x \left\{\; e\; J_{\rm
electron}\cdot A_{\rm EM}
- {1\over 8\pi}\left(E^2-B^2\right)_{\rm EM}\; \right\}\, ,
\eeq
where $K_{\rm electron}$ stands for the kinetic energy of the electron, and
where the
following two terms represent the interaction potential energy due
to the electron current $J$ and
electromagnetic field $A_{\rm EM}$, and the
Lagrangian for the free electromagnetic field,
in terms of the electric and magnetic fields,
themselves determined by derivatives of $A_{\rm EM}$. In
one or another form,
this expression can be found in textbooks for classical electromagnetism.

In these broad terms, the classical Lagrangian for chromodynamics looks much
the same,
\beqa
{\rm L}_{\rm QCD} &=& K_{\rm quark} - \int d^3x \left\{\; g\; J_{\rm
quark}\cdot A_{\rm QCD} -
{1\over 8\pi}\left(E^2-B^2\right)_{\rm QCD}\; \right\}\, ,
\eeqa
as the sum of quark kinetic, quark-chromodynamic field interaction, and pure
chromodynamic field terms.
In a slightly more elaborate but still
schematic form, we may represent ${\rm L}_{\rm QCD}$
 in terms of the fields themselves:
\beqa
{\rm L}_{\rm QCD}
&=& \int d^3x\ \left \{ \sum_q\, \bar{q}(x)\;\partial\; q(x)\; \right\}
\nonumber\\
&\ & \hspace{5mm} - g\; \int d^3x\ \left \{\sum_q\, \bar {q}(x)\; q(x)\;
A_{\rm QCD}\; \right\}
\nonumber\\
&\ & \hspace{10mm}
+ \int d^3x\ \left \{ (\partial A_{\rm QCD})^2 - g(\partial A_{\rm QCD})\;
A_{\rm QCD}^2
-g^2 A_{\rm QCD}^4\, \right\}\ ,
\label{qcdL}
\end{eqnarray}
where $q,\ \bar{q}$ represent spinor fields (quarks),
and $A_{\rm QCD}$ vector fields
(gluons) and $\partial$ represents space-time derivatives.
When fully expanded,
${\rm L}_{\rm QED}$ looks much the same in terms of
electron and photon fields, but without the cubic and quartic terms in $A$.

Even in this simplified form, Eq.\ (\ref{qcdL}) illustrates the basic principle
at the heart of the standard model, gauge symmetry.  We demand
the unobservability of the ``gauge''
transformation $q'(x)=\exp[i\alpha(x)]q(x)$,
even for a position-dependent (local) phase change $\alpha(x)$.
For QCD, the spinor field $q(x)$ is a vector in an
internal, color space, and $\alpha(x)$ is a matrix.
Any such transformation manifestly changes the first, kinetic
term in (\ref{qcdL}), as soon as the phase is space-time
dependent.  In a gauge theory, however, this change is cancelled by
a corresponding change in the interaction, $J\cdot A$, term,
when the vector field is modified in a corresponding manner:
$A_\mu'(x)=\exp[i\alpha(x)]A_\mu(x)\exp[-i\alpha(x)]
+(i/g)(\partial_\mu \exp[i\alpha(x)])\exp[-i\alpha(x)]$.
When the action for the vector field is built
along the lines of the final terms in Eq.\ (\ref{qcdL}), the Lagrangian
as a whole can be made invariant under the combined transformations.
This binding-up of the symmetry properties of spinor
and vector fields is encoded into every aspect of the standard
model.

The next step on the road to quantization is to transform the Lagrangian into
a Hamiltonian,
\beq
L \rightarrow H=H^{(0)}+gV_1+g^2V_2\, ,
\label{interactingH}
\eeq
where
$H^{(0)}$ is quadratic in the fields. In the perturbative approach,
we begin
by ``solving" $H^{(0)}$ that is,
  by identifying its eigenstates;
they will be free electrons and photons in QED, and free quarks and gluons in
QCD.
As we shall see, each term in $V=gV_1+g^2V_2$ defines an elementary process
that mixes free states.
Perturbation theory computes mixing as a power series in $g$,
an approximation to the true states and processes of the theory.

\subsection{From $H^{(0)}$ to Free-Particle States}

Let's construct the energy eigenstates of $H^{(0)}$. The natural degrees of
freedom are the spatial
Fourier transforms of the fields themselves, time-dependent functions
characterized
by wave numbers $\vec k$:
\beq
q(x),\ A^\mu(x)\, \rightarrow \tilde{q}(\vec{k},x^0),\
\tilde{A}^\mu(\vec{k},x^0)\, ,
\label{dof}
\eeq
with $x^0\equiv ct$ a measure of time in units of length.
We use $H^{(0)}$ as a starting point in the perturbative analysis of the
full Hamiltonian,
$H=H^{(0)}(q,A_\m)+V(q,A_\mu)$. The analysis is quite general, depending only
in the details.
By construction, $H^{(0)}$ is {\it quadratic} in
each of the degrees of freedom in (\ref{dof}). In the classical free theory,
this means that the fields obey a linear  equation, which
implies that solutions obey a principle of superposition: the sum of two
solutions is also a solution. In the wave number space of Eq.\
(\ref{dof}), the system always simplifies to an independent
harmonic oscillator equation of motion for each $\vec k$, with solutions
\beq
\tilde{q}(\vec{k},x^0) = b(\vec{k}) {\em e}^{-i\omega(\vec k)\, t}\, ,
\label{planewave}
\eeq
where $\omega(\vec k)$ is the frequency associated with wave vector $\vec k$,
and $b(\vec k)$ is
the amplitude of the corresponding
wave. Superposition implies that each $b(\vec k)$ is
independent.

The quantization of such plane waves is particularly
simple: each amplitude $b(\vec k)$ is quantized in the manner of a harmonic
oscillator.
The quantum system resides in states
characterized by quantized $|b(\vec k)|^2$
for each $\vec k$. These ``free" states may be written as
\beq
|m>=|\{\vec k_i\},\{\vec q_j\}>\, ,
\label{freestates}
\eeq
where each $\vec k_i$ denotes a quantum
excitation of wave vector $\vec k_i$ for the
spinor field, and $\vec q_j$ for the vector field.
The energies of the states (neglecting zero-point energies) are
\beq
H^{(0)}|m>=\hbar\; \left (
\sum_i\omega_i(\vec{k})+\sum_j\omega_j(\vec{q})\right )|m>
\equiv S_m|m>\, ,
\label{energies}
\eeq
with \begin{eqnarray}
\omega(\vec k)&=& \sqrt{\vec k^2c^2 +M^2c^4/\hbar^2}
={1\over \hbar}\sqrt{\vec p^2c^2+M^2c^4}
\nonumber\\
\omega(\vec q) &=&|\vec{q}| ={1\over \hbar} E(\vec q)\, ,
\end{eqnarray}
where $M$ is the mass of the spinor particle.
The second relation in Eq.\ (\ref{energies}) defines $S_m$.
The states $|m>$, being exact eigenstates of the Hamiltonian, are {\it
stationary}, that is, the
wave numbers describe the states for all time.  Another way of saying this is
that there is no scattering.  This is the
quantum version of the principle of superposition for the
equation of motion
of the classical field theory.  A typical state is illustrated
in Fig.\ \ref{treetopt}a, in which the horizontal direction represents time,
and two waves, the straight one representing an ``electron" and the
curved one a ``photon", pass through each other without interacting.

\subsection{The Interaction Mixes the Free States}

Scattering, interpreted as a change in the wave numbers of excitations
in the system, is associated with the potential terms $V$ in
the interacting Hamiltonian of Eq.\ (\ref{interactingH}).  To see why,
we only need to solve the Schr\"odinger equation in the
interacting theory:
\beq
i\hbar\, {\partial \over \partial t}|\psi(t)>= \left(
H^{(0)}+V\right)|\psi(t)>\, ,
\label{interactingSE}
\eeq
with a free-state boundary condition,
\beq
|\psi(-\infty)>=|m_0>\, .
\label{freeBC}
\eeq
The physical content of the boundary condition is intuitively clear;
it corresponds to the preparation of
  an experiment in which isolated
particles are arranged to collide.  This is what any accelerator
does.

In the perturbative solution to the interacting Schr\"odinger equation,
we assume that solutions to the free equation
are complete,
\beq
\sum_m |m><m| = 1\, .
\label{completeness}
\eeq
Introducing the notation,
\beq
V_{j,i} \equiv \langle m_j|V|m_i\rangle\, ,
\label{Vnotation}
\eeq
we may expand a general solution $|\psi(t)>$ in terms of the
states of the free ($V=0$) theory,
\beq
|\psi(t)>= \sum_m\; |m><m|\psi(t)>\, .
\label{expand}
\eeq
We readily verify that the following infinite
expansion for the matrix elements in (\ref{expand})
constitutes a solution to Eq.\ (\ref{interactingSE}),
\beqa
<m_n|\psi(t)> &=& \sum_{n=0}^\infty\; \sum_{m_1\dots m_n}\, {\rm
e}^{-iS_nt/\hbar}\;
(-i/\hbar)^n\, V_{n,n-1}\; V_{n-1,n-2}\times\dots \times V_{1,0} \, \nonumber
\\
&\ & \hspace{3mm} \times
\int_{-\infty}^t d\tau_n\; e^{-i(S_{n-1}-S_n)\tau_n/\hbar}
\int_{-\infty}^{\tau_n}d\tau_{n-1}e^{-i(S_{n-2}-S_{n-1})\tau_{n-1}/\hbar}
\nonumber
\\
&\ & \hspace{6mm} \times
\dots \times \int_{-\infty}^{\tau_2} d\tau_1 e^{-i(S_0-S_1)\tau_1/\hbar}\, ,
\label{ipsoln}
\eeqa
where the phases $S_m$ are defined by the free theory, through Eq.\
(\ref{energies}) above.

Although the expression for the matrix elements in
Eq.\ (\ref{ipsoln}) may seem a little complicated, it
has a nice interpretation.  Each term in the sums $\sum_{m_1\dots m_n}$
defines the evolution of the system, from an initial free state $|m_0>$
at $t=-\infty$ to
the observed state $|m_n>$ at time $t$, through a sequence of
free states  $|m_i>$, in which the system resides for a time
$\tau_{i+1}-\tau_{i}$.
At each step, the transition between states is governed by a
matrix element in the free
theory, $V_{m_{i+1},m_i}$.  The intermediate states need not be equal in energy
to the initial state, as reflected in the phases that
oscillate according to energy differences and intermediate times.  On
the other hand, if the potential is translation invariant, the
$V$'s will conserve momentum, by being proportional to delta functions of the
form
$\delta^3 (\vec p_{m_{i+1}}-\vec p_{m_i})$.

\begin{figure}[t]
\begin{center}
\hspace*{-7mm}
\epsfig{file=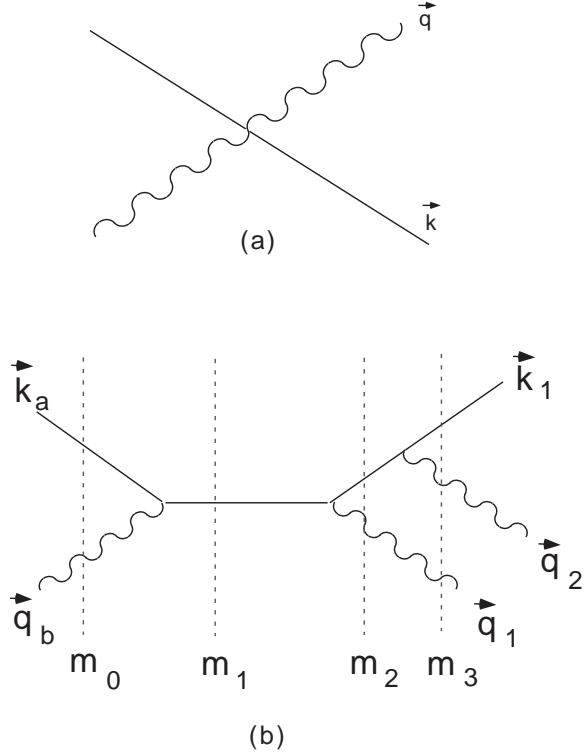,width=8cm}
\end{center}
\caption{(a) Representation of a state in the
free theory. (b) Mixing of states in the
interacting theory.}
\label{treetopt}
\end{figure}

An example is shown in Fig. \ref{treetopt}b,
in which an ``electron-photon"
system, with initial state $m_0$,
passes through two intermediate states,
$m_1$ and $m_2$, ending up in a state, $m_3$, with an electron
and
two photons.
Each vertex in this ``time-ordered diagram'' represents one of the
matrix elements $V_{j,i}$.  The lines meeting at any vertex
match the fields in the corresponding term in the potential.
Thus, for QED, with potential $e\bar q q A_{\rm QED}$, each
vertex connects two spinor and one vector line.  QCD
has, in addition, three-vector ($gA_{\rm QCD}^3$) and 
four-vector ($g^2A_{\rm QCD}^4$) vertices.

\subsection{Time-ordered Perturbation Theory}

To put Eq.\ (\ref{ipsoln}) into a
  more standard form, we simply do the time
integrals,
which are trivial. For every $\tau_i$, $i<n$, they give
$i/(S_{m_{i}}-S_0)$.  If we
take the limit, $t\to\infty$ for the last time,  $\tau_n$, its
integral gives
$2\pi\delta(S_n-S_0)$, which enforces energy conservation over long enough
times.

\begin{figure}[h]
\begin{center}
\hspace*{-7mm}
\vspace*{3mm}
\epsfig{file=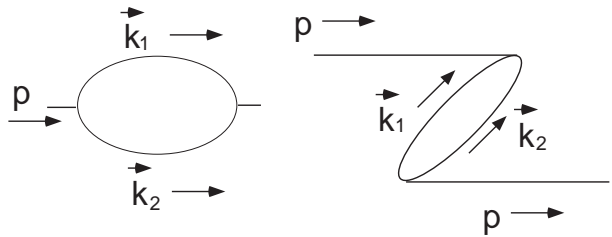,width=8cm}
\end{center}
\caption{Time ordered diagrams for a self-energy.}
\label{onelooptopt}
\end{figure}

The result is time-ordered
(sometimes called ``old-fashioned") perturbation theory.
\footnote{From here on, we shall go over to units in which
all momenta and masses are specified in dimensions of
$\rm (length)^{-1}$.  Thus, for an energy $E$ or mass $M$,
we define $p_0=E/\hbar c$, and $m=Mc/\hbar$, respectively.
We shall also drop the arrow over wave numbers.}
Let us define
$\lim_{t\to\infty}<m|\psi(t)>\equiv\Gamma(p)$,
where $p$ denotes collectively the wave numbers of particles in the initial
state $m_0$ and
the final state $m_n$.  The general form for $\Gamma$, suppressing the energy
conservation
delta function, is
\footnote{The term $i\epsilon$ in the denominators defines the
integrals at $t=-\infty$ in Eq.\ (15) by attributing a
small imaginary part to the incoming energy $S_0$.}
\beq
\Gamma(p)=
\prod_{{\rm loops}\ i}\int{d^3\ell_i\over (2\p)^3}\prod_{{\rm lines}\
j}{1\over 2\omega_j(k_j)}
\prod_{{\rm states}\ a} {1\over E_a-S_a+i\epsilon}\, N(p,\ell_i)\, .
\label{topt}
\eeq
The function $N$ collects overall
factors and polynomials in momenta from the product of the
matrix elements $V$.
These depend
on the spin and other quantum numbers of the fields.
The sum over states  in Eq.\ (\ref{topt}) is
\beq
\sum_{{\rm states}}=
\prod_{{\rm loops}\ i}\int{d^3\ell_i\over (2\p)^3}\prod_{{\rm lines}\
j}{1\over 2\omega_j(k_j)}\, .
\nonumber
\eeq
Here the independent, or ``loop" momenta, $\ell_i$ are the momenta 
left over after
all momentum conservation delta functions in the matrix elements have been
employed.
An example is shown in  Fig.\ \ref{onelooptopt},
which is a ``self-energy",
the amplitude for a particle,
of initial energy $p_0$, to evolve into itself, at second order in a
potential
that links three fields together.
This amplitutde is
\beqa
&&\Gamma_1 +\Gamma_2 =
\\
&& \hspace{3mm}
\sum_{\rm 2-particle\ states}\left ( {1\over
p_0-\omega(k_1)-\omega(k_2)+i\epsilon}
+ {1\over- p_0-\omega(k_1)-\omega(k_2) +i\epsilon} \right )\, ,
\nonumber
\eeqa
where we have suppressed all overall factors and delta functions.  The two
terms correspond
to the two time orderings.  In the first of the orderings, the potential
transforms the
particle into a two-particle state, whose energy is
$\omega(k_1)+\omega(k_2)$.  In the
second, the potential actually creates three particles out of the vacuum, so
that
the intermediate state contains four particles, and has energy
$2p_0+\omega(k_1)+\omega(k_2)$,
as reflected in the different ``energy denominator" in this case.

Because energies are not Lorentz invariant, the contributions of individual
time-orderings
to an amplitude are also not invariant.  Nevertheless, their sum is.
This overall invariance is made manifest by the formalism of Feynman diagrams,
each of which summarizes the complete set of time orderings
of the same topology.  Thus, the two time-ordered diagrams of Fig.\
\ref{onelooptopt} may
be represented by a single invariant diagram, without relative ordering of
its vertices.
The price for doing this is to introduce a new energy integral for each loop,
and
to replace energy denominators by covariant propagators, one for each line.
For the self-energy above, we find
\beqa
G(p) \equiv
\Gamma_1 +\Gamma_2 =
  \int {d^4k \over (2\pi)^4}\ {1\over k^2 -m^2+i\epsilon}\; {1\over (p-k)^2-m^2
+i\epsilon}\, ,
\eeqa
or more generally, summing over all orders in an arbitrary diagram,
\beq
G(p)\equiv\sum_{\rm {\sl t}-orders}{\Gamma(p)} = \prod_{{\rm loops}\, i}
\int {d^4\ell_i\over (2\pi)^4} \prod_{{\rm lines}\, j}{i\over
k_j^2(p,\ell_i)-m_j^2+i\epsilon} \tilde{N}(p,\ell_i)\, .
\eeq
The factors $\tilde N$ are again associated with the quantum number content
of the
fields.

This brings our development of the formalism of quantum field theory to a
conclusion.
Luckily, to understand the basic issues in the phenomenology
of quantum field theory, it is not necessary to delve into too many of the
theory-dependent details.  Most of our discussion
will rely
on the description of scattering amplitudes as sums over all possible paths
through
intermediate states.

\section{Using Perturbation Theory: Searches}

It is not too misleading to characterize
the aim of high energy physics as the identification of an underlying
Hamiltonian, or action, that governs the time development of the universe.
We think that we know parts of that Hamiltonian, at least its effective
form at the energy scales that we can investigate directly.  In experiments,
we try to learn about the parts that we do not yet know.
The essential strategy is simple: predict the outcomes of scattering
experiments based upon the known Hamiltonian, and look for deviations.
As deviations are found, identify a modified Hamiltonian that accounts
for them, and which can be tested through its predictions.  Of course,
the experiments that we choose to test our current theory will depend on
how we perceive its successes and its shortcomings.

\subsection{The Cycle of Tests}

The cycle of testing
to which we have just referred proceeds through the computation of
quantum-mechanical scattering amplitudes, and hence probabilities.
A broad summary is as follows.

\begin{itemize}

\item{}  Lists of states available to us are the eigenstates of the
contemporary $H^{(0)}$.

\item{} The rules by which states mix are given by the contemporary $V$.

\item{} It is useful to classify states according to
symmetry generators $S_i$ (momenta and charges) that commute with
the contemporary $H=H^{(0)}+V$.  States are identified by their
conserved charges, as well as their momenta: $S_i\, |m,\{s_i\}\rangle = s_i\,
|m,\{s_i\}\rangle$.

\end{itemize}

In this context, {\it new physics} may require us to make changes.

\begin{itemize}

\item{} We may have to
add new states to our list, and hence reformulate $H^{(0)}$.  Old states
may turn out to be composites of new states, and $H^{(0)}$ may well
simplify in the process.

\item{} We may have to add new rules for changing states, and hence modify the
potential, perhaps adding a new term, $\delta V$.

\item{} If $[\delta V, S_i]\ne 0$, some
  symmetry $S_i$ of our original Hamiltonian
is violated, which implies
that the corresponding charge $s_i$ may not be conserved.  On the other
hand, new states and interactions may possess new symmetries and conserved
quantum numbers.

\end{itemize}

\subsection{Favorite Tests}
\label{testssec}

Practitioners of high energy
physics have developed, over time, a toolbox
of generic experiments, sensitive to the capabilities of quantum mechanical
scattering.

\begin{figure}
\begin{center}
\hspace*{-7mm}
\vspace*{3mm}
\epsfig{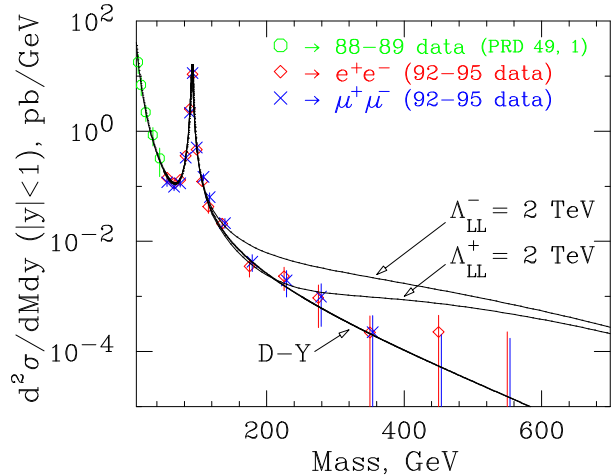}
\end{center}
\caption{High mass lepton pairs at the Tevatron.}
\label{dyfig}
\end{figure}

\begin{figure}
\begin{center}
\hspace*{-7mm}
\vspace*{3mm}
\epsfig{file=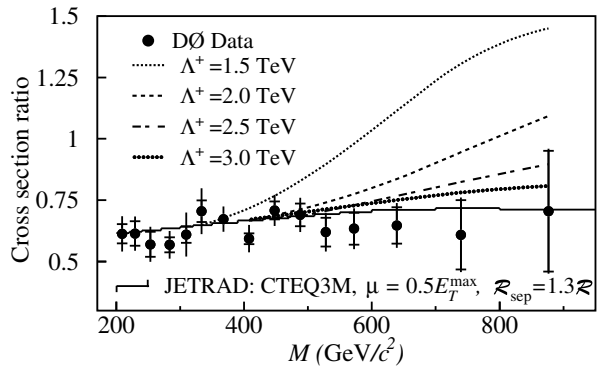,width=8cm}
\end{center}
\caption{Compositeness limits from the high
mass jets at the Tevatron.}
\label{contactfig}
\end{figure}

\begin{figure}
\begin{center}
\hspace*{-7mm}
\vspace*{3mm}
\epsfig{file=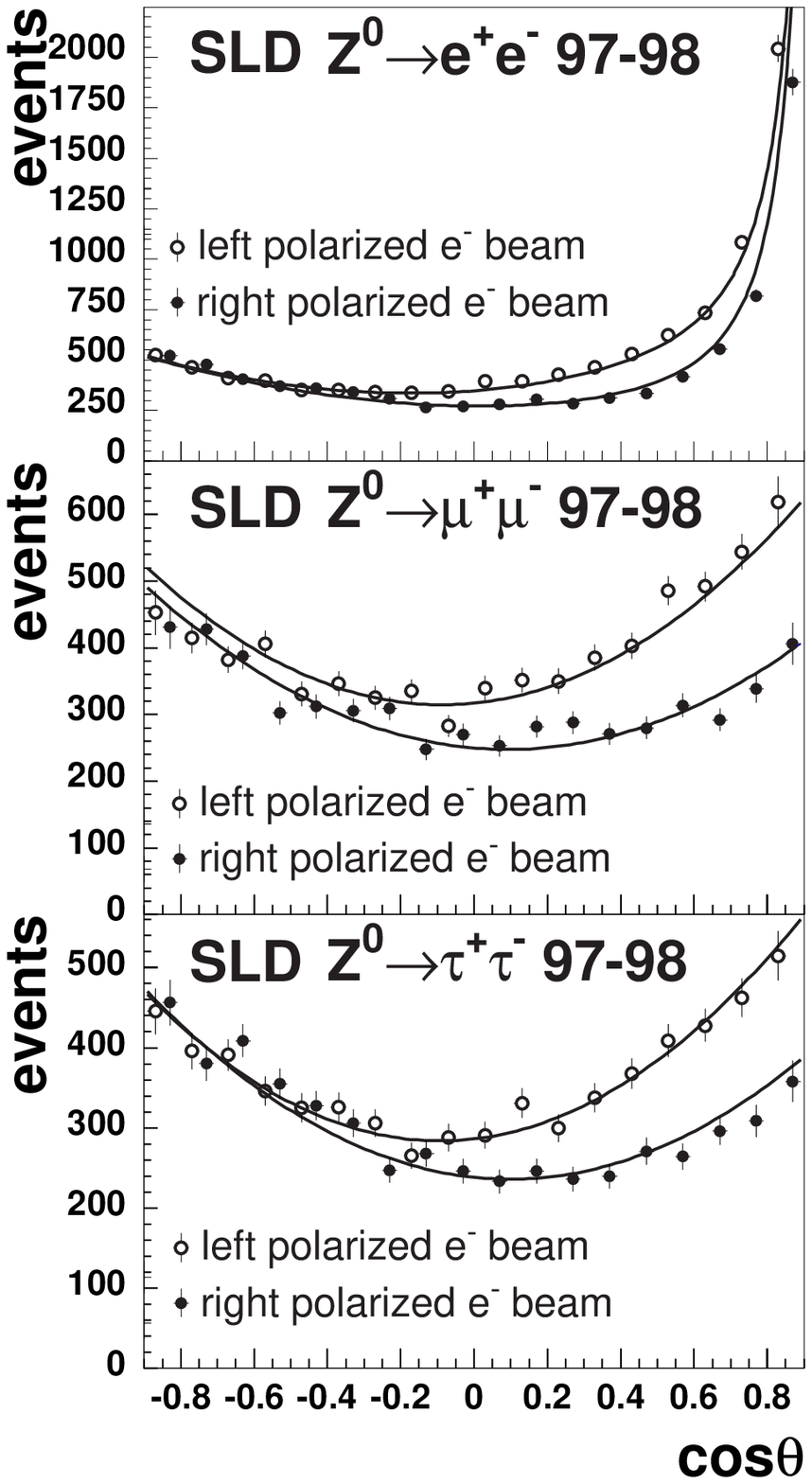,width=5cm}
\end{center}
\caption{Forward-backward distributions
of lepton pairs.
Symmetry under parity would
require that the two
curves  be the same in each case.}
\label{fbasym}
\end{figure}

\begin{itemize}

\item{}
As the energy increases, we are able to produce states of higher mass.
When the energy is just right for the production of a particle through
a hitherto unseen term in the Lagrangian by a combination of
known particles, a state with the new particle is formed. This matching of
energies
expresses itself through an increase in the amplitude,
and hence the probability, as the energy deficit decreases.
An example is shown in Fig.\ \ref{dyfig}, which is the
cross section (the probability normalized to the density of
colliding particles), for the production of
particle-antiparticle pairs of
leptons through the annihilation of pairs of quarks.
Generally, the cross section decreases with energy, but
when the energy of the quark-antiquark pair matches the
mass of the Z boson, at around 90 GeV, the cross section
increases dramatically.  This data was taken by the CDF
collaboration to scan for new physics at yet higher
energies.\ \cite{cdf}  As we can see, there is no sign yet.

The LEP II accelerator at CERN has searched for
the Higgs scalar particle, H, through the reaction
$e^++ e^- \rightarrow {\rm Z+H}$.  The Higgs is
the only remaining state, and
field, that is part of the standard model Hamiltonian,
but which has not yet been observed
definitively.  In the standard model, we can readily
compute the scattering amplitude for this process as a function of the
mass of the Higgs scalar.  If the Higgs particle is found in this way,
its mass -- a previously unknown parameter of the Hamiltonian --
will be determined, and the Hamiltonian of the standard model dramatically
verified.   Other popular searches are for particles whose
presence is implied by supersymmetry, or by ``extra" space
dimensions.

\item{}  We may infer the presence of very heavy states even when we
cannot produce them directly, if they couple
to known fields through terms in the Hamiltonian.
Very heavy particles, which appear in virtual states that live for only a very
short time may produce nonstandard {\it contact terms}, through which standard
fields appear to interact.  Thus, in high energy experiments, it is
important to look for new interactions between  leptons and
quarks, or quarks and quarks:
\beq
\delta L_{\ell q}
  = {1\over \Lambda^2}\ \sum_{ij}(\bar e_i\Gamma e_i)(\bar q_j\Gamma
q_j)\
\hspace{3mm}
\delta L_{q q}
  = {1\over \Lambda^2}\ \sum_{ij}(\bar q_i\Gamma q_i)(\bar q_j\Gamma
q_j)
\, ,
\eeq
where $\Gamma$ represents a appropriate matrices.
Here $\Lambda$ is actually an energy denominator, associated with
states of high mass.
Because $\Lambda$ is much larger than the initial state energy,
this denominator is effectively constant.
Such a term would produce a
very different angular distribution than the standard model interaction,
which always involves virtual states with one of the known
vector bosons: the gluon,
photon, W and Z.  In Fig.\ \ref{contactfig}, the effects
of contact terms, interpreted as signals of quark compositeness,
are compared to
jet cross sections (see below) observed by the D0 detector at
Fermilab \cite{d0}.  The curves show the kind of limits that can be
put on contact terms today, typically requiring the scale $\Lambda$
to be a few TeV, roughly one thousand times the rest energy of the proton.

\item{} Other favorite tests involve decays
that are forbidden, or suppressed, in the standard  model.
Examples include a muon changing to an electron, and a bottom quark to
a strange quark,
\beq
\hspace{-60mm}
\mu \rightarrow e + \gamma\, , \quad\quad
{\rm b}\rightarrow {\rm s} + \gamma\, ,
\eeq
where the first is absent, and the latter rare, in the
standard model.  The existence of the former would imply
physics ``beyond" the standard model.  The latter, which
has been seen, requires a detailed evaluation of standard
model predictions on rates and final states.
Current limits \cite{pdg} for $\mu \rightarrow e+\gamma$ are of the
order of 10$^{-11}$; b decay to s occurs at the level
of 10$^{-4}$.

\item{} Symmetry violation.  The archetype of
symmetry violation searches
are the famous parity-violation experiments of the fifties,
which showed that
the Hamiltonian of the weak interactions did not respect
right/left reflection.
Although this was a clue to the form of the relevant
Hamiltonian, much more time, data and thought was necessary to
discover the weak sector of the standard model.
An illustration of how a symmetry violation manifests itself
is shown in Fig.\ \ref{fbasym},
the angular distribution of lepton pairs in the decay of
a Z at the SLD detector at SLAC, for different orientations
of the spins of the incoming electrons
that produced the Z.\ \cite{sld}  If parity were
respected in this experiment, the curves
would be identical for different polarizations.

\end{itemize}

Experiments that search for new, heavy particles require
imparting energies to conventional particles that are high
enough to cross, or at least approach, the relevant energy thresholds.
The menagerie of high energy machines consists of a few very
large animals.  Their species are: machines for
electron-positron collisions, as at CERN (LEP II) and SLAC (SLC),
electron-proton
collisions, as at DESY (HERA), and proton-antiproton, or proton-proton
collisions, as at Fermilab (Tevatron) and CERN (LHC).
Many variations are possible, but the aim of searches is
to find new physics through new particles and new effects in
scattering final states.  The search for rare decays and
symmetry violation requies, in the main, rate rather than
energy. Thus, for example, ``factories'' for the production
of B mesons (CESR at Cornell,
Belle at KEK and BaBar at SLAC) are tuned
to an appropriate energy range where they
are accumulating vast quantities of events.

\subsection{The Program: Ideal and Reality.}

The ideal program for the determination of fundamental laws of nature in the
language
of quantum field theory is, of course, difficult
to realize.
On the experimental side, nature has evidently arranged energy scales
in such a way as to make their exploration nothing if not
challenging.  Theoretical
difficulties
are in many ways analogous to those of Newtonian gravity.  The
very richness of any theoretical structure
adequate to describe a varied class of phenomena requires
new ideas
and methods for its implemention.  Thus, although expressions like
Eq.\ (\ref{topt}) were derived very early after the invention of quantum
field theory, it was a long time before anyone was able to do the sums
over states
in any but the simplest cases, and even
today there are many restrictions
to our abilities in this
regard.  In the remainder of this talk, I'd like to describe
what some of these problems are, and indicate some of the methods developed
to deal with them.

The most common, and fundamental, problem with the calculation of scattering
amplitudes
according to Eq.\ (\ref{topt}) is that the sums over intermediate states
almost never converge without further input.
In quantum field theory, the sums over free particle states
are integrals over possible wave numbers.
As it stands, we integrate over all wave numbers, including those that
are arbitrarily large, corresponding to very high energy modes, and those
that are very small, corresponding to low energy modes.  By the usual
correspondence of resolution to wavelength, the former are
sensitive to the short-distance dynamics of the theory, the latter
to its long-distance behavior.

Problems  arise at short distances because, if $k$ is a typical\
wave number, transition amplitudes often increase with $k$, roughly
as  $V_{n,n-1} \sim k^a$, with $a={\sum (spins)}$, where ``spin" refers to the
intrinsic angular momentum carried by each  of the fields.  This
behavior leads  to nonconvergence in most theories in four dimensions
(even those with spinless particles).  In particular, it limits the
set of theories that have a reasonable perturbative interpretation at all.

Even when a theory can be controlled at short distances,
integrals over long wavelengths can refuse to converge,
a situation
referred to as an infrared divergence.  Infrared divergences
indicate strong sensitivity to long-time behavior, and indeed, they
are related to the difficulties encountered in the study of
planetary motion
in Newtonian gravity, which is sensitive to small, but cumulative,
effects in the solar system.

Almost all scattering amplitudes suffer from either or
both of these problems at some order in the potential, $V$.
The question is therefore how to get any useful information
at all out of such an ill-defined scheme!
A partial, but workable, solution is to separate dynamics at different scales.
That is, when the contributions from long or short wavelengths
are not well predicted by our theory, we attempt to organize
our ignorance into a few parameters, or even a few functions,
that we take as given -- determined by experiment.  We reduce
the theory to calculations over only those wave numbers that
are not too large, and not too small.
In the following section, we discuss how divergences at short
distances are handled through renormalization, and at
long distances, through infrared
safety and factorization.  The limitations of these
approaches are many, including a general lack of mathematical
definition, approximations that
are at best asymptotic series, and
corrections whose coefficients cannot be bounded rigorously.
It is far from obvious
at the outset that such a program can work, but in fact it
has been remarkably successful.

\begin{figure}
\begin{center}
\hspace*{-7mm}
\vspace*{3mm}
\epsfig{file=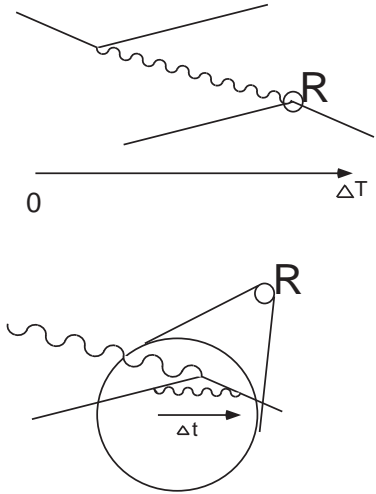,width=5cm}
\end{center}
\caption{Gluon exchange and vertex correction.}
\label{scales}
\end{figure}

\section{Separating Scales}

\subsection{Short Times and Renormalization}

Problems with the sum over states at short distances
are illustrated by Fig.\ \ref{scales} in quantum electrodynamics.
Here we show a low-order
contribution to  the development  of a state with
two electrons over some fixed time $\Delta T$, during which a virtual
intermediate state with an electron and a photon appears
for a while.  At the end, we revert to a two-electron state, but
in the process momentum may be transferred, which we
interpret as the quantum-mechanical origin of the force between
electrons.  In fact, the sum over states shown in the figure
can be carried out in QED, and gives a result that has
been known for a long time.  We may decide,  however, to take
a closer look at the process, in particular, at the region
in time denoted by the circle $R$ in the figure.
The lower part of Fig.\ \ref{scales}
shows a blowup of $R$, and on this scale we see that the
absorption of the photon at the second vertex on top
happened while the receiving electron was itself in a virtual state
involving another photon.  That state lasted for a time
$\Delta t\ll \Delta T$,  and we might well have missed it
on the time-scale at the top of Fig.\ \ref{scales}.

The practical problem in QED, QCD and any other four dimensional quantum
field theory is that when $\Delta t\rightarrow 0$, all of these
diagrams diverge, from their sums over very high energy states.
By the correspondence between short times and high energies,
these are the very short-lived
states.

\begin{figure}
\begin{center}
\hspace*{-7mm}
\vspace*{3mm}
\epsfig{file=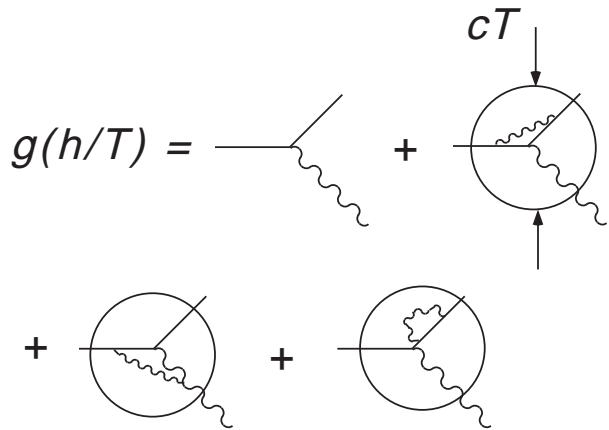,width=8cm}
\end{center}
\caption{Lowest-order
diagrams that contribute to the running coupling in QCD.}
\label{RngCpg}
\end{figure}

The solution to this problem, known as renormalization, is
really rather straightforward.  We simply replace all interactions
like the bottom of Fig.\ \ref{scales},
on time
scales less than or equal to {\it any} fixed time $T$, by a known
function, called the  {\it running coupling $g(\hbar/T)$},
where $\hbar/T$ is the energy characteristic of frequency $1/T$.
We would like to include in the running coupling the sum of the
diagrams shown in Fig.\ \ref{RngCpg}, including {\it all}
virtual states that are within time $T$,
or distance $cT$, of the
electron-photon vertex.  Equivalently, we
would like to sum over
all intermediate states with energies greater that $\hbar/T$.
Even though the sum does not converge, we can still identify
the $T$-dependence of the running coupling, because an infinitesimal
change in $T$ only involves an infinitesimal change in energy,
and therefore depends on only a tiny part of
the whole sum.  Therefore, we can
derive a  differential equation for $g\left(\mu\right)$,
of the form
\beqa
{\partial g(\mu) \over \partial \ln \mu}=
\beta \left(g(\mu)\right) &=& - {g^3(\mu) \over
16\pi^2}b_0+ {\cal O}\left(g^5(\mu)\right)\, ,
\label{beta}
\eeqa
where a direct calculation from the diagrams of Fig.\ \ref{RngCpg} gives
\beqa
b_0 &=& 11- 2n_f/3 \quad \quad (\rm QCD)
\label{RCevol}
\eeqa
with $n_f$  the number of different relevant kinds (flavors)
of quarks.  Roughly speaking, a quark is relevant when its mass is
no larger than $\mu$, so that pairs of quarks may appear in
virtual states on the time scale $T=\hbar/\mu$.

Once we have an approximate differential equation for the running coupling,
we can solve for it, and we find, 
\beq
g^2(\hbar/T) = {16\pi^2 \over b_0 \ln(\hbar^2/T^2\Lambda_{\rm QCD}^2c^2)}\, ,
\label{gsquare}
\eeq
where $\Lambda_{\rm QCD}$ is a mass that specifies a boundary condition --
which can be taken as the
value of $g(\mu)$ at any fixed value of $\mu$.  $\Lambda_{\rm QCD}$ is
sometimes referred to as the ``QCD scale",
the scale at which the strong
interactions become strong.  Its actual value is
not accessible to computation within QCD; presumably it is
set by the physics at very short times,
where the states are no longer adequately
described by this theory.  The magic of renormalization is
that, whatever those states are, at low
energies in QCD they express themselves
only in the value of $\Lambda_{\rm QCD}$.
This {\it decoupling} of physics at very short
times is what makes it possible for us to systematize
our knowledge of QCD and the rest of
the standard model self-consistently.
It also helps us to understand why it is sometimes
so difficult to see beyond the successes of
the standard model.

\subsection{The case of QCD}

The solution, Eq.\ (\ref{gsquare}) for the QCD running coupling has
features which are unique to
that theory.  In particular, when measured
at short times, $T\to 0$, the coupling is weak, and even vanishes
over very short time scales, or equivalently at very short wavelengths,
or again equivalently, at very high energies.  This property is
called {\it asymptotic freedom}.     Correspondingly, however,
when measured over long times, or at low energies,
the coupling becomes strong, and indeed diverges at times
of order $1/\Lambda_{\rm QCD}$.  This behavior corresponds nicely
to the paradoxical picture of QCD that
emerges from high energy experiments.  Over short
distances, where large momenta can be transferred,
the theory acts as though it were well described in terms
of the states of $H^{(0)}_{\rm QCD}$, that is, in
terms of quarks and gluons.
We shall see how shortly.
  But if we wait long enough
(or go far enough into the past),
the quarks and gluons {\it always} and without fail
conspire to form color-neutral states
$|m_{final}\rangle$ ($|m_{initial}\rangle$),
consisting of mesons and baryons with $q\bar q$ and $qqq$ quantum numbers,
occasionally with quark-less ``glueballs''.
This is  known as {\it confinement}.  Qualitatively, all
this is consistent with the behavior of the running coupling
in (\ref{gsquare}), but it is not immediately obvious
how to combine these features into a usable theory.
Here again we  turn to a separation of scales.
In this case, we shall learn how to separate long times from short.

\subsection{Large Times and Physical Pictures}

To separate long from short times in matrix elements, we need to see
how the former contribute to time-ordered perturbation theory.
Certain very general features are easy of identify.   For this
purpose, we return to Eq.\ (\ref{ipsoln}), and see that
the time integrals are all of the same kind,
\beqa
&& \int_{-\infty}^t d\tau_n\; e^{-i(S_{n-1}-S_n)\tau_n}
\int_{-\infty}^{\tau_n}d\tau_{n-1}
e^{-i(S_{n-2}-S_{n-1})\tau_{n-1}} \dots
\nonumber\\
&& \hspace{20mm}
\times \int_{-\infty}^{\tau_2}
d\tau_1 e^{-i(S_0-S_1)\tau_1}\, .
\label{onetimeintegral}
\eeqa
In fact, it is not so easy to generate sensitivity to long-time
intervals in these expressions, because all the time dependence is in
exponentials, and oscillating $t$-dependence suppresses large times as
the integrals cancel over each oscillation.
The exception to this principle is
at points of stationary phase in Eq.\ (\ref{onetimeintegral}).  Examining
the exponents shows that the
total phase has a curious interpretation, which
may be represented as:
\begin{eqnarray}
{\rm PHASE}\,&=& \sum_{{\rm states}\ m=1}^n S_m (\tau_m-\tau_{m-1}) \nonumber
\\
&\ & \nonumber \\
&=& \sum_{{\rm states}\ m=1}^n \left ( \sum_{{\rm particle}\ j\ {\rm in}\ m}
\omega({p}_j) \right ) (\tau_m-\tau_{m-1}) \nonumber \\ &\ & \nonumber \\
&=& \quad {\rm FREE-PARTICLE\ ACTION}\, .
\end{eqnarray}
Thus, the phase may be thought of as the sum of all the {\it classical}
action accumulated in the intermediate states, consisting of free particles.
In other words, stationary phase corresponds to stationary classical action.
But stationary action corresponds to physical motion in the classical
mechanics of particles.  Thus, surprisingly, sensitivity to long times
requires that the succession of states in the computation of the
scattering amplitude have a description as a succession of free particles
propagating between points in space-time.
When this is {\it not} the case, the amplitude is dominated by
canceling phases, and there is no sensitivity to long-time dynamics.
Quantities that are independent of long-time dynamics are sometimes called
{\it infrared safe}.

\subsection{Infrared Safe Cross Sections}

Are there any infrared safe observables?  By definition, they
must be independent of how quarks are confined, or any other
long-time properties of QCD.  In fact, there are many, but they
always involve {\it inclusive} measurements.  They are
seen most directly in $\rm e^+e^-$ collisions, as at the LEP
machine at CERN.  The simplest example is
probably the total cross section
for an electron-positron pair to
annihilate, that is to mix with a state which consists of
a single photon, which subsequently decays to
hadrons, $\sigma_{\rm tot}^{({\rm e}^+{\rm e}^-)}$.
\footnote{The photon state through which the system passes is usually
referred to as an ``off-shell" photon, because its energy $E$
is much larger than $pc$, its momentum times the speed of light.
  $E=pc$ is the ``mass shell" relation for the (massless)
photon.}  Alternately, the electron-positron pair may be
just energetic enough to produce an on-shell Z particle,
whose decay rate, $\Gamma_{\rm Z}$, can then be measured.  The
rate is just a  normalized
probability per unit time (the larger the probability the shorter the
lifetime).

\begin{figure}
\begin{center}
\hspace*{-7mm}
\vspace*{3mm}
\epsfig{file=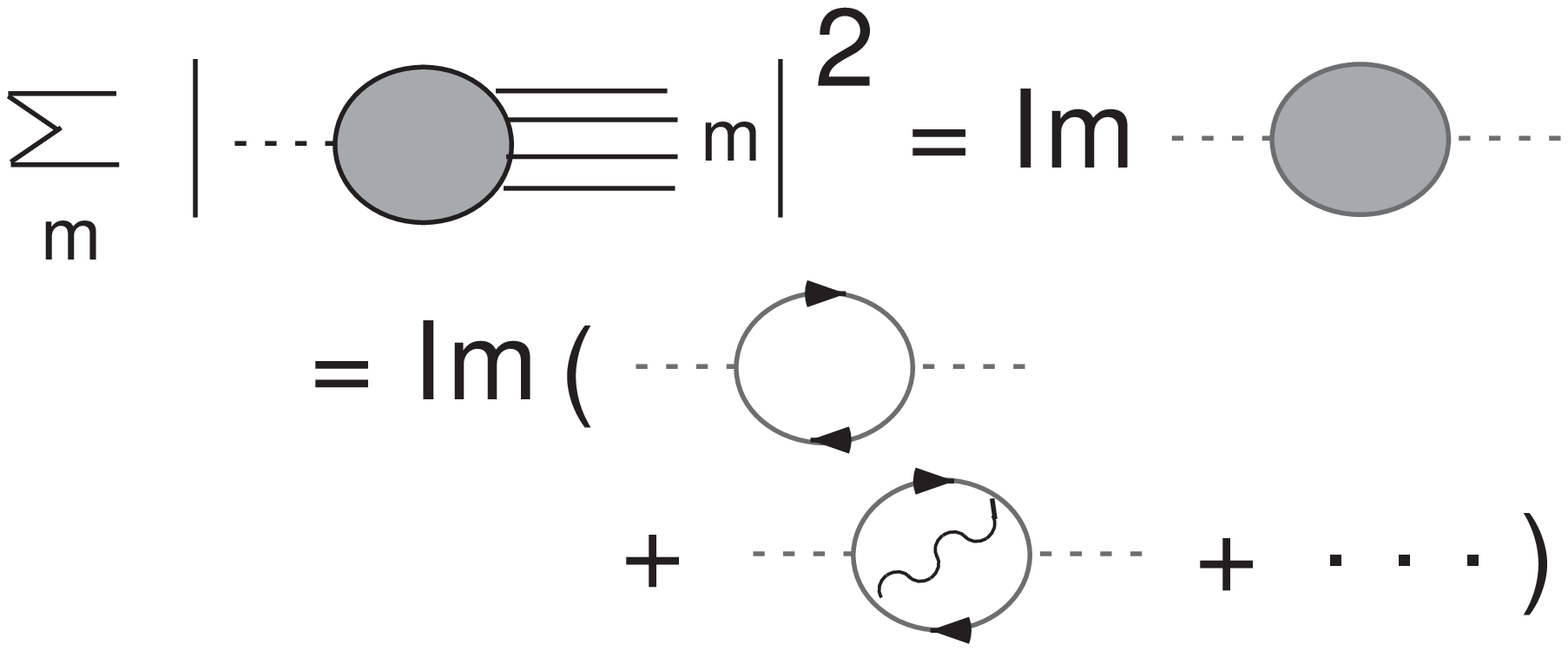,width=10cm}
\end{center}
\caption{The optical theorem for Z decay.}
\label{Unitarity}
\end{figure}

The first step in showing that $\Gamma_{\rm Z}$ is infrared
safe is to invoke the conservation of probability, in the form
of the
optical theorem of quantum mechanics, which states that the
total decay probability
per unit time is
(up to some constants which we ignore)
the imaginary part of the amplitude for {\it forward} scattering.
This is shown schematically in Fig.\ \ref{Unitarity}, whose
left-hand side represents the sum over all possible final states
for the decay of a Z particle (the dashed line).  By the optical
theorem, then,
\beq
\Gamma_{\rm Z} \sim {\rm Im}\, \Pi_{\rm Z}(Q^2=m_{\rm Z}^2)\, ,
\label{optical}
\eeq
where $\Pi_{\rm Z}(Q^2)$
is the imaginary part of the forward-scattering
amplitude, for a Z to change into a Z by passing through any and
all intermediate states that the field theory allows.

\begin{figure}
\begin{center}
\hspace*{-7mm}
\vspace*{3mm}
\epsfig{file=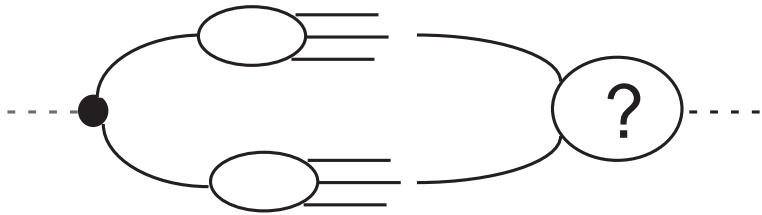,width=10cm}
\end{center}
\caption{Lack of physical processes for Z decay}
\label{Zdecay}
\end{figure}

Now let's think about  the forward scattering amplitude.
We ask whether there are any physical pictures for $\Pi_{\rm Z}(Q^2)$,
in which the Z decays at a point into particles which
travel freely, rearrange themselves at points into other
states, but eventually come back together to form the Z.
There are no such physical pictures, because after the
Z decays the particles it produces will be moving rapidly
in opposite directions, as illustrated in Fig.\ \ref{Zdecay},
and there is no way for them to reassemble the Z
by free propagation.  Like Humpty-dumpty, the Z cannot
be put back together again: q.e.d.

This may seem a bit abstract, but if we go ahead and calculate
the Z decay rate using Eq.\ (\ref{topt}), the infrared safety
that we have just proved guarantees that we will find a completely
finite answer.  That is, the sums over intermediate states
converge at long wavelengths, and we find
a finite expansion in $\alpha_s(M_{\rm Z})$:
\beq
\Gamma_{\rm Z}=\Gamma_{\rm Z}^{\rm (EW)}\, \left(1+\sum_{n=1}^\infty
c_n\alpha_s^n(M_{\rm Z})\right)
\,  ,
\eeq
where $\Gamma_{\rm Z}^{\rm (EW)}$ is the decay rate
with no QCD interactions at all (EW stands for electroweak).
In the expansion, $\alpha_s\equiv g^2/4\pi$ is
the QCD ``fine-structure constant''.
Because all the integrals are finite, we can choose the scale of our
running coupling to be $M_{\rm Z}$.  Then, neglecting the masses
of the quarks, the coefficients $c_n$ are just numbers,
because there are no further dimensionless ratios on which they
can depend.
For example,
$c_1=1/\pi$.

With one more technical observation, we can extend
these considerations to a much larger class of observables:
the {\it jet} cross sections.  A jet in $\rm e^+e^-$ annihilation
or Z decay
is just a collection of particles moving in more-or-less the same
direction.  The complete details of the identities and momentum
spectra of particles within a jet do not turn out to be
infrared safe, but the probability distribution for a jet
with a specified {\it total} energy moving in a specified
direction, does turn out to be infrared safe.

Suppose we ask for the probability,
not just that the Z should decay, but that it
should decay in such a way that all
but a small fraction, $\varepsilon$, of its energy flows into two 
oppositely-directed
cones of angular size $\delta$:
\begin{equation}
\sigma_{\rm 2J}= \sigma(E_{{\rm cones}\,
(\delta)}\ge (1-\varepsilon)m_{\rm Z})\, .
\end{equation}
This would be a very unlikely arrangement if particles were distributed
randomly in the final state.
Imagine, however, a Hamiltonian for QCD in which particles are
produced {\it only} in the directions of these cones.
Such a Hamiltonian would not be Lorentz, or even rotationally
invariant, but it could be a bona-fide Hamiltonian anyway,
and we could apply all the reasoning that we used for the
full Hamiltonian to this truncated one.
If we use unitarity for such a Hamiltonian, the sum over
all of its available states will be infrared safe, just as
for the full Hamiltonian, and should be computable in
perturbation theory, with a result
of the form
\begin{equation}
\sigma_{\rm 2J}= \sigma_0\, \sum_n c_n(\delta)\, \alpha_s^n\, ,
\label{2jirsafe}
\end{equation}
with $\sigma_0$ a convenient normalization
and with coefficients $c_n$ that depend on
the opening angles $\delta$.
The crucial observation now is that the theory with this
truncated Hamiltonian has, for this particular cross
section, exactly the same set of points of stationary
phase in the sum over its intermediate states, and therefore
the same long time behavior, as the {\it full} theory.
This is because for a jet final state the
physical pictures of evolution in the full theory {\it only}
involve the decay and recombination of particles within
the jet cones, or the emission of very soft particles.  Intermediate
states with energetic particles moving in directions outside
of the jet cones are never at points of stationary
phase, because once outside the cones, the particles never
encounter other particles that could scatter them
back in.  Thus, the truncated and full theories only differ
by short-distance, infrared safe, corrections in
perturbation theory.  As a result, the form (\ref{2jirsafe})
holds for full QCD perturbation theory.
This reasoning can be extended to a large class of
jet-related cross sections in
${\rm e^+}{\rm e}^-$ annihilation.
To be specific, we can identify a flexible
criterion: any cross section that is insensitive to collinear
rearrangements and to emission of soft gluons
is infrared safe.

If we measure a cross section for jets defined in
this fashion, we have a way of ``seeing'' quarks
and gluons.  Of course, there is no unique jet
definition; each event contains, according to quantum
mechanics, a sum over its possible histories.
For infrared safe cross sections, however,
we can compute corrections to the dominant history,
and predict the outcomes of variant definitions
of jet events.
In this way, we can design experiments to
study the true short-distance behavior of a theory,
even though at short times the very degrees of
freedom are different from those that we detect
directly.

\subsection{Factorization and Deep-inelastic Scattering}

With jet cross sections, we have opened a window to
the short distances of quantum chromodynamics.  The
set of such infrared safe cross sections
is rather limited, however, and does not include any
experiment where we start out with strongly interacting
particles (hadrons).  Why not?  When there are hadrons
in the initial state, the
strong interactions have been going on since time
immemorial;  the quarks in a random proton may well have
been interacting with each other since the big bang.
Thus, there is no hope that  these
long-time interactions cancel out.  They were there
to begin with, and their presence actually defines the cross
section.  Still, we need not abandon hope.
Since every proton is the same, initial-state interactions
  may appear as overall factors in  suitably-defined cross
sections.  The trick is to identify
the appropriate class of reactions.
In a sense, this is easy.  We should study the same sets
of final states as in $\rm e^+e^-$ annihilation --
inclusive cross sections, but always involving the large
momentum transfers that imply short-time interactions.
The archetype of these reactions is deep-inelastic
scattering, from which, in fact, arose the first direct evidence
for quarks.

The deep-inelastic scattering of an electron and a proton
(or nuclear) target involves reactions of the sort
\beq
e(k)+N(p) \rightarrow e(k')+X(p+q)\, ,
\eeq
where the invariant momentum transfer, $q^2\equiv -Q^2 = -(k'-k)^2$,
and energy transfer in the target rest frame,
$\nu\equiv (p\cdot q)/m_N$
are both large.  The first condition ensures that the scattering takes
place over a short distance, the second that it takes place
over a short time.   In this case, the process proceeds
at the quantum mechanical level through the exchange
of a photon between the electron and a quark in the target,
similarly to the top of Fig.\ \ref{scales}.

Because the electromagnetic interaction is much weaker
than the strong, most of the complexity is in the
succession of states of the proton, which fluctuate
much more rapidly than those of the electron.
As a result, we can work to ``lowest order'' in electromagnetism,
and keep only states in which a single photon is exchanged.

In a typical succession of states that describes
this process in field theory, the electron and proton
fluctuate into an electron plus a photon for the former,
and a collection of quarks and gluons for the latter.
In this, rather complex, intermediate state, one of the quarks
absorbs the photon, to produce a state which is almost the
same as the original virtual state of the proton,
differing only in the momentum of
the ``active'' quark, the one that received the photon, which
is now moving in a new direction, with an energy that
is much larger than the rest mass of the proton.
The basic reaction is the same as in the transition
from state $m_0$ to state $m_1$ in Fig.\ 1b.

The experiment is particularly simple, because to
measure the momentum transfer, $q$, it is only necessary
to look at the electron in the final state.
The inclusive cross section for all deep-inelastic reactions
with the same momentum and energy transfer,
has much in common with the total electron-positron annihilation
cross section, and can be treated in much the same way.
Summing over all final states by means of the optical
theorem, we again arrive at a forward-scattering cross section.
Now, instead of a decay, however, the forward scattering
describes a process in which the intermediate photon
is absorbed by the constituents of the proton, and is
then reemitted, into the same momentum state.  Just
as in the annihilation case, the absorption and reemission cannot be
separated by states in which the scattered, active quark
propagates too far.  If it did, it would
recede too far from the remnants of the proton for the
proton to reform in the forward scattering.
The only
physical pictures available for forward scattering are
those in which the photon is absorbed and emitted at points
that are separated by a light-like distance within
the proton.
In the center-of-mass system of the proton
and virtual photon, Lorentz contraction makes even this is a short distance.
As a result, all of the  effects associated with
the evolution of final states, such as the probabilities
of producing jets,
are calculable, just as in electron-positron annihilation.
The same is not true of interactions in the initial state.
These are, however, quite independent of the
deep-inelastic scattering itself.  These
arguments suggest that the short-distance interactions,
including those that determine the jet structure of the
final state, are quantum-mechanically incoherent
with the initial state interactions that bind the
hadronic target.  This incoherence has powerful implications.

First of all, because the proton target has been around
for a long time, the  contribution of the initial, struck
quark, ${\cal Q}$, to the energy deficits of states before the
interaction must be relatively small.  For this to be the
case, the momentum of that quark
  should be given approximately
  by $\xi p$, with $0\le \xi \le 1$.  A quark with a large
transverse momentum, or with a fraction $\xi$ outside
this range, would automatically produce a state
that is very short-lived, and which would therefore
contribute a perturbatively-calculable correction.
At the same time,
the incoherence of the scattering process
with the binding of the proton means that
the scattering can be described by transitions between an
initial state, $\gamma+{\cal Q}$, to a set of final states, the
simplest of which is a single quark ${\cal Q}$,
\beqa
\gamma(q) + {\cal Q}(\xi p) \rightarrow && {\cal Q}(\xi p+q)\, ,
  \nonumber\\
\rightarrow && {\cal Q}(\xi p+q-k) + {\cal G}(k)\, \dots\, ,
\label{process}
\eeqa
with $\cal G$ a gluon.
The requirement that the final-state jets produced
by the struck quark be physical is $(\xi p+q)^2>0$,
which, neglecting masses, leads to the restriction
\beqa
\xi \ge {Q^2 \over 2p\cdot q} \equiv x\, ,
\label{xdef}
\eeqa
where  $x$ is called the ``Bjorken scaling variable''.

In technical terms, these considerations are summarized
by a set of very general relations.
For  simplicity, let's stick to the fully inclusive
cross section, which depends only on the momentum transfer $q$
and the momenta of the initial particles,
$k$ for the electron, and $p$ for the target nucleon.
The ``one-photon'' approximation for the electrodynamics
part of the virtual states allows us
to write the cross section as a product
of ``leptonic'' and ``hadronic'' functions:
\beq
\sigma_{eN} = L_e(k,k')\; \times\; W_N(q,p)\, .
\eeq
The  incoherence between initial and final
state strong interactions in  the scattering process
is expressed in what is known as a factorization formula
for the hadronic part,
\beqa
W_N(q,p) &=& \sum_{a={\cal Q},\bar{\cal Q},{\cal G}}\,
\int_x^1 dy\; C_a\left({q^2\over \mu^2},{x\over y},\alpha_s(\mu)\right)\;
f_{a/N}(y,\mu)\, .
\label{W_N}
\eeqa
The sum is over all parton types.  At higher orders
in QCD, even a gluon may scatter from the photon,
by transforming itself into a virtual
state with a quark-antiquark pair.
The functions $C_a$ are called coefficient
functions.  They describe the dynamics of the
possible scatterings in Eq.\
(\ref{process}) at
short times, and are
calculable in perturbation theory.  The
$f_{a/N}$ are parton distribution functions, which
interpolate between hadrons and  partons.
The scale $\mu$ in these functions
is the inverse of the largest lifetime of
a state that we allow into $C_a$:
  $\mu=1/T_{\rm max}$.  Quantum mechanical incoherence
implies that these functions appear as a factorized
product in the calculation of $W_N$.
  At present, we can't calculate the parton distributions,
but we can measure them,
by fitting Eq.\ (\ref{W_N}) to data over a range of
momentum transfers, and by observing weak boson as
well as photon exchange.

The parton distribution
functions may be expressed as
expectation values in states of the target,
for example, for quark ${\cal Q}$ in nucleon $N$,
\beqa
f_{{\cal Q}/N}(\xi,\mu)
=\int d\lambda\; {\rm e}^{-i\lambda \xi p\cdot n}\; \langle
N(p)|
\bar {\cal Q}(\lambda n)
  \Gamma_{\cal Q} {\cal Q}(0)|N(p)\rangle\, ,
\eeqa
where ${\cal Q}$ is the corresponding quark field
operator, and where
$\Gamma_{\cal Q}$ is a matrix that
projects out a number operator
for quarks of momentum fraction $\xi$.
The vector $n^\mu$ represents the light-like velocity
of the struck quark.
The factorization
scale $\mu$ enters this expression as the scale of
renormalization in the perturbative calculation of the
matrix element.

A striking consequence of asymptotic
freedom follows from
  Eq.\ (\ref{W_N}).  As the momentum transfer
$Q$ increases, we may take the factorization
scale $\mu$ to be large, of order $Q$.
But at any $x$ and $Q$, $W_N$ must be independent of our
choice of $\mu$:
\beq
\mu{d W_N(p,q)\over d\mu}=0\, .
\eeq
This means that the $\mu$-dependence of the parton distributions
must be compensated by that of the coefficient functions.  But the
coefficient functions are calculable in perturbation theory.
Hence, so must be the $\mu$-dependence of the parton distributions:
\beq
\mu\, {df_{a/N}(\xi,\mu^2)\over d\mu}
= \sum_{b={\cal Q},\bar {\cal Q},{\cal G}}\int_\xi^1 dz\;
P_{ab}(\xi/z,\alpha_s(\mu))\; f_{b/N}(z,\mu)\, .
\eeq
The only dimensional scale upon
which the  ``evolution kernels" $P_{ab}$
may depend is $\mu$ in  $\alpha_s(\mu)$, because $\mu$ is the
only such variable held in common by $C$ and $f$.  This means
we can take parton distributions determined from one set of
data, and apply them to predict scattering at much higher,
or lower, momentum transfers, to any scale for which the
running coupling remains small.

As the factorization scale  $\mu$ increases,
the coupling becomes weaker and
weaker, and the importance of
extra gluons in the final state in Eq.\ (\ref{process})
becomes less and less, until only the
simple {\it quark + photon $\rightarrow$ quark} process
remains.  Eventually,
$W_N$ becomes barely dependent on $Q$,
a property known as {\it scaling},
in which dependence on the scaling variable
$x$ of Eq.\ (\ref{xdef}) is all that remains.
Scaling was
for QCD what elliptic orbits were for
Newtonian gravity, a dramatic explanation
of a striking, but previously unaccountable,
observation.

Beyond this, the factorization formalism makes it possible to
predict cross sections for jets and particle production
at arbitrary length scales, not only in electron-nucleon
scattering, but also in nucleon-nucleon scattering.
As a practical matter, this enables us to
make predictions for
higher energies, and to search for new physics,
perhaps some new particle $F$ of mass $Q$.
In hadronic collisions,
the factorized inclusive cross section
for any such ``$F$-production'',
at center-of-mass energy $E_{\rm cm}$, involves two momentum fractions,
one for each hadron,
\beqa
Q^4\, {d\sigma_{AB\to F(Q)+X}\over dQ^2} &=& \sum_{a,b}
\int_0^1 {d\xi \over \xi}\;
\int_0^1 {d\eta \over \eta}\;
\hat\sigma^{\rm (PT)}_{ab\rightarrow F(Q)+X}\left({Q^2\over \xi\eta E_{\rm
cm}^2},\mu,\alpha_s(\mu)\right)\;
\nonumber\\
&\ & \hspace{10mm} \times
f_{a/A}(\xi,\mu)\, f_{b/B}(\eta,\mu)\, ,
\eeqa
with the same parton distributions $f$ as in deep-inelastic scattering.
As for $C_a$ in Eq.\ (\ref{W_N}),
$\hat \sigma_{ab}^{\rm (PT)}$ is perturbatively calculable.
It is on the basis of formulas such as this, that the Tevatron, and
subsequently
the LHC, will search for new physics, by constantly comparing data to
predictions based on the standard model and its extensions, through a
cycle of tests as described above.  Already volumes
of calculations and predictions exist, each implementing a proposal
for what will be found in the short-distance cross sections
$\hat \sigma_{ab}^{\rm (PT)}$.
The curves in Figs.\ 3 and 4 were calculated in this way.
Nature will decide which, if any, matches
the data.

\section{Conclusions}

In broad terms,
the physics of fundamental processes today continues
as it did three centuries ago, though tests of our
existing understanding of matter and forces.  These
tests have a dual role, first in the elaboration of
our current models, and second, in a search for their
limitations.

We are far from a full command of our present
quantum field theories.  Our ability
to separate scales, however, enables us to probe
different sectors of a theory on paper,
on the computer screen and in the
laboratory, through renormalization and factorization.
For QCD, in particular, each hadronic event displays the
imprint of dynamics at all scales.
In the observation of its dynamics we cannot escape
the dualities of  weak and strong, fundamental
and effective.  The mathematician-physicists
of the eighteenth century developed tools to
apply the law of gravitation to the motions of the
moon and planets, from the book of astronomical observation.
So are we attempting to apply
modern field theory to hadronic scattering, and
to read the quantum mechanical history
of chromodynamics in the alphabet of its final states.

The search for new physics, to reveal a
structure that will explain the standard model and
its apparent complexities in terms of something (hopefully)
simpler,  may only require the energy necessary
to narrow an energy deficit,
give a  heavy virtual state just a little
longer lifetime, and produce the kinds of
signals described in Sec.\ \ref{testssec}.
Because of the reach of hadronic colliders,
in particular the Tevatron and the LHC, these will most
likely be the key to unlock the next energy scale.
If they do, a new chapter in our
understanding of fundamental process will
begin, with a new world of physics and  mathematics
to explore.

\section*{Acknowledgements}

I would like to thank the organizers of the {\it International
Conference on Fundamental Science:
Mathematics and Theoretical Physics} for the invitation to
participate in a stimulating meeting, and for their
hospitality during my visit.  In particular, I
thank Prof.\ C.-H.\ Oh for help and encouragement.  I would
also like to thank Brookhaven National Laboratory for
its hospitality during the preparation of the manuscript.
This work was supported in part by the National Science
Foundation, grant PHY9722101.


\section*{References}

\begin{thebibliography}{99}

\bibitem{Hankins} T.L.\ Hankins, {\it Jean d'Alembert} (Oxford Univ.\ Press,
Oxford, 1990).

\bibitem{giudice} G.F.\ Giudice, at the 19th International
Symposium on Lepton and Photon Interactions at High Energy
(LP99), eConf {\bf C990809}, 440 (2000); e-Print Archive hep-ph/9912279.

\bibitem{cteq} CTEQ Collaboration,
{\it Rev.\ Mod.\ Phys.}\ {\bf 67}, 157 (1995).

\bibitem{tasigs} G.\ Sterman, in {\it QCD and Beyond},
proceedings of the Theoretical Advanced
Study Institute (TASI 1995),  ed.\ D.E.\ Soper (World Scientific,
Singapore, 1996), e-Print Archive hep-ph/96063312.

\bibitem{tasids} D.E.\ Soper, lectures at the Theoretical Advanced
Study Institute (TASI 2000), e-Print Archive hep/ph-0011256.

\bibitem{cdf} CDF Collaboration (F.\ Abe
{\it et al.}), {\it Phys.\ Rev.\ Lett.}\ {\bf 79},
2198 (1997).

\bibitem{d0} D0 Collaboration
(B.\ Abbott {\it et al.}), {\it Phys.\ Rev.\ Lett.}\ {\bf 82},
2457 (1999), e-Print Archive hep-ex/9807014.

\bibitem{pdg} D.E.\ Groom {\it et al.}\ (Particle Data Group),
{\it Eur.\ Phys.\ J.}\ {\bf C15}, 1 (2000).

\bibitem{sld} SLD Collaboration, e-Print Archive hep-ex/0010015.


\end{thebibliography}
\end{document}